\documentclass[prl,twocolumn,showpacs, superscriptaddress]{revtex4-1}
\usepackage[margin=1in]{geometry}
\usepackage{hyperref}
\hypersetup{
  colorlinks   = true, %Colours links instead of ugly boxes
  urlcolor     = blue, %Colour for external hyperlinks
  linkcolor    = blue, %Colour of internal links
  citecolor   = red %Colour of citations
}
\usepackage{graphicx}
\usepackage{amsmath}
\renewcommand{\vec}{\mathbf}

\begin{document}

\title{Staggered Dynamics in Antiferromagnets by Collective Coordinates}

\author{Erlend G. Tveten}
\author{Alireza Qaiumzadeh}
\affiliation{Department of Physics, Norwegian University of Science and Technology, NO-7491 Trondheim, Norway}
\author{O. A. Tretiakov}
\affiliation{Institute for Materials Research, Tohoku University, Sendai 980-8577, Japan}
\affiliation{Department of Physics \& Astronomy, Texas A\&M University, College Station, Texas 77843-4242, USA}
\author{Arne Brataas}
\affiliation{Department of Physics, Norwegian University of Science and Technology, NO-7491 Trondheim, Norway}
\date{\today}

\begin{abstract}
Antiferromagnets can be used to store and manipulate spin information, but the coupled dynamics of the staggered field and the magnetization are very complex.
We present a theory which is conceptually much simpler and which uses collective coordinates to describe staggered field dynamics in antiferromagnetic textures. 
The theory includes effects from dissipation, external magnetic fields, as well as reactive and dissipative current-induced torques.
We conclude that, at low frequencies and amplitudes, currents induce collective motion by means of dissipative rather than reactive torques.
The dynamics of a one-dimensional domain wall, pinned at 90$^{\circ}$ at its ends, are described as a driven harmonic oscillator with a natural frequency inversely proportional to the length of the texture.
\end{abstract}


\maketitle

New developments have created opportunities for using antiferromagnets (AFMs) as active components in spintronics devices~\cite{MacDonald13082011}. 
AFMs are ordered spin systems which lack a macroscopic magnetization in equilibrium because neighboring spins compensate each other. 
Analogous to ferromagnets, in AFMs domain walls can be engineered~\cite{logan:192405}, the anisotropic tunneling magnetoresistance (AMR) is substantial~\cite{Park:2011rt,*PhysRevB.79.134423, *PhysRevB.81.212409}, spin-wave logic gates can be useful~\cite{Rovillain:2010fk}, and the order parameter can be switched ultra-fast by light~\cite{Kimel:2009fk}. 
Additionally, AFMs have no stray fields, and high-temperature AFM semiconductors can be realized~\cite{PhysRevB.83.035321}, enabling control of the carrier concentration governing all transport properties.

In magnetic materials, currents induce torques on the magnetic moments~\cite{Brataas:2012fk}. 
In ferromagnets, these torques can be used to switch the magnetization, induce steady state precession in magnetic oscillator circuits, or move domain walls.
Theoretical~\cite{PhysRevB.73.214426,*PhysRevB.75.014433, *PhysRevB.75.174428, *Gomonay2008, *PhysRevLett.100.196801, *PhysRevB.81.144427,*PhysRevB.85.134446} and experimental~\cite{PhysRevLett.98.116603, *PhysRevLett.98.117206, *PhysRevLett.99.046602} results indicate that current-induced torque effects are present in AFMs as well, and that these effects are of the same order of magnitude as in ferromagnets.   
However, several aspects are fundamentally different.
For instance, the dynamics in AFMs are described by coupled equations of the staggered field and the (out-of-equilibrium) magnetization. Current-induced torques affect these variables differently.

In AFMs, the staggered field may spatially vary and is influenced by external magnetic fields and currents.
Traditionally, understanding the complex behavior of the temporal- and spatial-dependent order parameter requires solving a set of coupled equations with many degrees of freedom.
In this Letter, we formulate a conceptually simpler theory of how external forces influence the staggered field and magnetization dynamics in AFMs in terms of a few collective coordinates.
Our description is based on the phenomenological theory of insulating AFMs~\citep{lifshitz1980}, extended to account for charge current flow~\cite{PhysRevLett.106.107206}, making the theory valid also for metallic and semiconducting AFMs.
It includes the effects of dissipation, external magnetic fields, and both reactive (adiabatic) and dissipative (non-adiabatic) current-induced torques in slowly varying inhomogeneous antiferromagnetic textures.

Consider a basic AFM lattice consisting of two magnetic sublattices, with magnetic moments $\vec{m}_{1}(\vec{r},t)$ and $\vec{m}_{2}(\vec{r},t)$, so that the total magnetization is $\vec{m}(\vec{r},t)=\vec{m}_{1}(\vec{r},t)+\vec{m}_{2}(\vec{r},t)$, and the antiferromagnetic order parameter is $\vec{l}(\vec{r},t)\equiv\vec{m}_{1}(\vec{r},t)-\vec{m}_{2}(\vec{r},t)$.
In the absence of magnetic fields and textures, the equilibrium magnetization vanishes and $\vec{l}(\vec{r},t)$ is finite and homogeneous.
Below, we consider the dynamics of the magnetization vector and the unit N\'{e}el vector $\vec{n}(\vec{r},t)=\vec{l}(\vec{r},t)/l(\vec{r},t)$.

To the lowest order in textures and magnetizations, the AFM free energy reads~\cite{lifshitz1980,PhysRevLett.106.107206}
\begin{equation}
\label{eq:freeenergy} U=\int d\vec{r}\left[
\frac{a}{2}\vec{m}^{2}+\frac{A}{2}\sum_{i=x,y,z}(\partial_{i}\vec{n})^{2}-\vec{H}\cdot\vec{m}\right],
\end{equation}
where $a$ and $A$ are the homogeneous and inhomogeneous exchange constants, respectively. 
$\vec{H}$ represents the external magnetic field. 
From the free energy (\ref{eq:freeenergy}) and the constraints $|\vec{n}|=1$ and $\vec{m}\cdot\vec{n}=0$, which are valid for temperatures well below the N\'{e}el temperature, we can construct the effective fields $\vec{f}_{n}=-\delta U/ \delta\vec{n}=A\vec{n}\times(\nabla^{2}\vec{n}\times\vec{n})-\vec{m(\vec{H\cdot\vec{n}})}$ and $\vec{f}_{m}=-\delta U/ \delta\vec{m}=-a\vec{m}+\vec{n\times(\vec{H\times\vec{n}})}$. 
In all our results, we may generalize the free energy (\ref{eq:freeenergy}) by adding anisotropy terms, e.g., easy-axis anisotropy $K_{z}n_{z}^{2}/2$.

Hals \textit{et.~al.}~\cite{PhysRevLett.106.107206} introduced phenomenological reactive (adiabatic) and dissipative (non-adiabatic) current-induced torque terms, as well as dissipation. With these additional terms, the equations of motion are
\begin{eqnarray}
\label{eq:currentmotion}
\dot{\vec{n}} & = & (\gamma\vec{f}_{m}-G_1\dot{\vec{m}})\times\vec{n}+\eta\gamma(\vec{J}\cdot\nabla)\vec{n},\\
\label{eq:currentmotion2} 
\dot{\vec{m}} & = & (\gamma\vec{f}_{n}-G_2\dot{\vec{n}}+\beta\gamma(\vec{J}\cdot\nabla)\vec{n})\times\vec{n}+T_{nl},
\end{eqnarray}
where $\gamma$ is the gyromagnetic ratio, $G_{1}$ and $G_{2}$ are phenomenological Gilbert damping parameters, and $\eta$ ($\beta$) parametrize the adiabatic (non-adiabatic) current-induced torque terms.
Throughout this paper, we disregard all non-linear terms that are contained in $T_{nl}$~\cite{PhysRevLett.106.107206}.
Eqs.~(\ref{eq:currentmotion}) and (\ref{eq:currentmotion2}) are the AFM analogs to the Landau-Lifshitz-Gilbert-Slonczewski equation for ferromagnets.
By combining these equations, the magnetization can be expressed in terms of the AFM order parameter, giving a closed equation for the staggered field vector $\vec{n}$ to the linear order in the out-of-equilibrium deviations $\vec{m}, \partial_t\vec{n}, \vec{J}$, and $\vec{H}$:
\begin{eqnarray}
\label{eq:neeldynamics} \frac{\ddot{\vec{n}}}{\tilde{\gamma}} & = &
-\vec{n}\times\dot{\vec{H}} + G_{1}\dot{\vec{f}}_{n}+(\eta
+G_{1}\beta) (\dot{\vec{J}}\cdot\nabla)\vec{n} \nonumber\\ & & +
a[\gamma\vec{f}_{n}-G_{2}\dot{\vec{n}}+\gamma\beta(\vec{J}\cdot\nabla)\vec{n}].
\end{eqnarray}
Here $\tilde{\gamma}\equiv\gamma/(1+G_{1}G_{2})$ is a modified effective gyromagnetic ratio in the presence of dissipation.
Eq.~(\ref{eq:neeldynamics}) is the starting point for deriving the collective coordinate equations of motion for AFMs.

In ferromagnets, magnetic textures are often rigid, so that only a few, \emph{soft} modes dominate the magnetization dynamics, as in the seminal work of Schryer and Walker on domain wall motion~\cite{schryer:5406}.
The evolution of these soft modes can be described by a finite set of \textit{collective coordinates}. 
This approach greatly simplifies the understanding of complex magnetization dynamics, making it possible to approximately describe the dynamics at low energies by considering only a few soft modes.

The collective coordinate approach has recently been applied to magnetization dynamics in ferromagnets~\cite{PhysRevLett.100.127204, *PhysRevB.78.134412}.
We now present how the equations of motion for the collective coordinates can be constructed for AFMs. 
We transform Eq.~(\ref{eq:neeldynamics}) by requiring the time dependence of the N\'{e}el field to be described by a set of collective coordinates $\{b_{i}(t)\}$: $\vec{n}(\vec{r},t)\equiv\vec{n}(\vec{r},\{b_{i}(t)\})$. 
The time derivative of the staggered field is then $\dot{\vec{n}}=\dot{b}_{i}\partial_{b_{i}}\vec{n}$.
Similarly, $\ddot{\vec{n}}=\ddot{b}_{i}\partial_{b_{i}}\vec{n}+\mathcal{O}(\dot{b}_{i}^{2})$, where the second term is disregarded in our linear response analysis since it is quadratic in the driving forces.

The dissipation is described in Eq.~(\ref{eq:neeldynamics}) via the terms $G_{1}\dot{\vec{f}}_{n}$ and $aG_{2}\dot{\vec{n}}$. 
The first term scales as $G_{1}A/(\lambda^{2} \tau)$, where $\lambda$ and $\tau$ are characteristic length and time scales of the staggered field texture. 
The second term scales as $aG_{2}/\tau$. 
In analyzing the relative strengths of these dissipative terms, we use the fact that the homogeneous and the inhomogeneous exchange constants are related through $a\sim A/(l^{2}d^{2})$~\cite{SovPhysUspBaryakhtar1985}, where $d$ is the lattice constant and we have introduced the AFM order parameter $\vec{l}$ above.
Dissipation in metallic ferromagnets is small since it arises from the spin-orbit interaction in combination with electron scattering~\cite{PhysRevB.84.054416}.
It is likely that similar mechanisms in AFMs are also weak, and that they have comparable effects on the staggered field and the magnetization:
$G_{1}l\approx G_{2}/l \ll 1$.
From this we can conclude that $\tilde{\gamma}\approx\gamma$ and that the second dissipative term, $aG_{2}\dot{\vec{n}}$, dominates in realistic systems, where the typical size of the texture $\lambda$ is such that $\lambda\gg d$.
Hence $G_{1}\dot{\vec{f}}_{n}$ can be safely disregarded in the equation of motion~(\ref{eq:neeldynamics}).

Our main result is the equations of motion for the soft modes:
\begin{equation}
\label{eq:eqmotion}
M^{ij}(\ddot{b}_{j}+\gamma aG_{2}\dot{b}_{j}) = F^{i}.
\end{equation}
This equation is derived by introducing the collective coordinates to Eq.~(\ref{eq:neeldynamics}), taking the scalar product with $\partial_{b_{j}}\vec{n}$, and integrating over the space.
The dynamics are equivalent to the classical motion of a massive particle subject to dissipation-induced friction and external forces.
This equation is model independent and can be used to determine the parameters of AFMs, e.g.~the Gilbert damping $G_2$ and the homogeneous exchange constant $a$, which are usually difficult to identify in experiments.

In Eq.~(\ref{eq:eqmotion}), $M^{ij}$ is the effective mass arising from the exchange interaction between the spins. 
The total force inducing motion of the collective coordinates, ${{F}}^i={{F}}^i_{{X}} + F^i_{{J}}+F^i_{{H}}$, is a sum of the exchange force, the current-induced force, and the external field force:
\begin{subequations}
\begin{eqnarray}
M^{ij}(\vec{b})&=&\frac{1}{a\gamma^2}\int dV\,\partial_{b_{i}}\vec{n}\cdot\partial_{b_{j}}\vec{n},\label{eq:collectivemass}\\
F^i_{{X}}(\vec{b}) &=& \int dV\,\partial_{b_{i}}\vec{n}\cdot\vec{f}_{n},\label{eq:collectiveforce}\\
F^i_{{J}}(\vec{b})&=&\int dV \Big[\beta \partial_{b_{i}}\vec{n}\cdot(\vec{J}\cdot\nabla)\vec{n}
\nonumber\\
&&+\frac{\eta+G_{1}\beta}{a\gamma} \partial_{b_{i}}\vec{n}\cdot(\dot{\vec{J}}\cdot\nabla)\vec{n}\Big],\label{eq:collectivecurrent}\\
F^i_{{H}}(\vec{b})&=&\frac{1}{a\gamma} \int dV\,\dot{\vec{H}}\cdot(\vec{n}\times \partial_{b_{i}}\vec{n}).\label{eq:collectivefield}
\end{eqnarray}
\end{subequations}
More generally, Eq.~(\ref{eq:collectiveforce}) can also be expressed as $F^i_{X}=\partial_{b_{i}} U$, to include the effective material-specific forces which act on the AFM through the exchange interaction and magnetic anisotropy.
Eq.~(\ref{eq:collectivecurrent}) includes the reactive and dissipative current-induced forces, both of which are important for the dynamics of the collective coordinates $b_{i}$.
Eq.~(\ref{eq:collectivefield}) represents the response to an external magnetic field.
Note that in the linear response regime, only time varying external magnetic fields affect the dynamics of the collective coordinates in AFMs, in contrast to the situation for ferromagnets~\cite{PhysRevLett.100.127204}, making the collective motion in AFMs more resistant to stray fields.

\begin{figure}[h]
        \centering
        \includegraphics[width=0.48\textwidth]{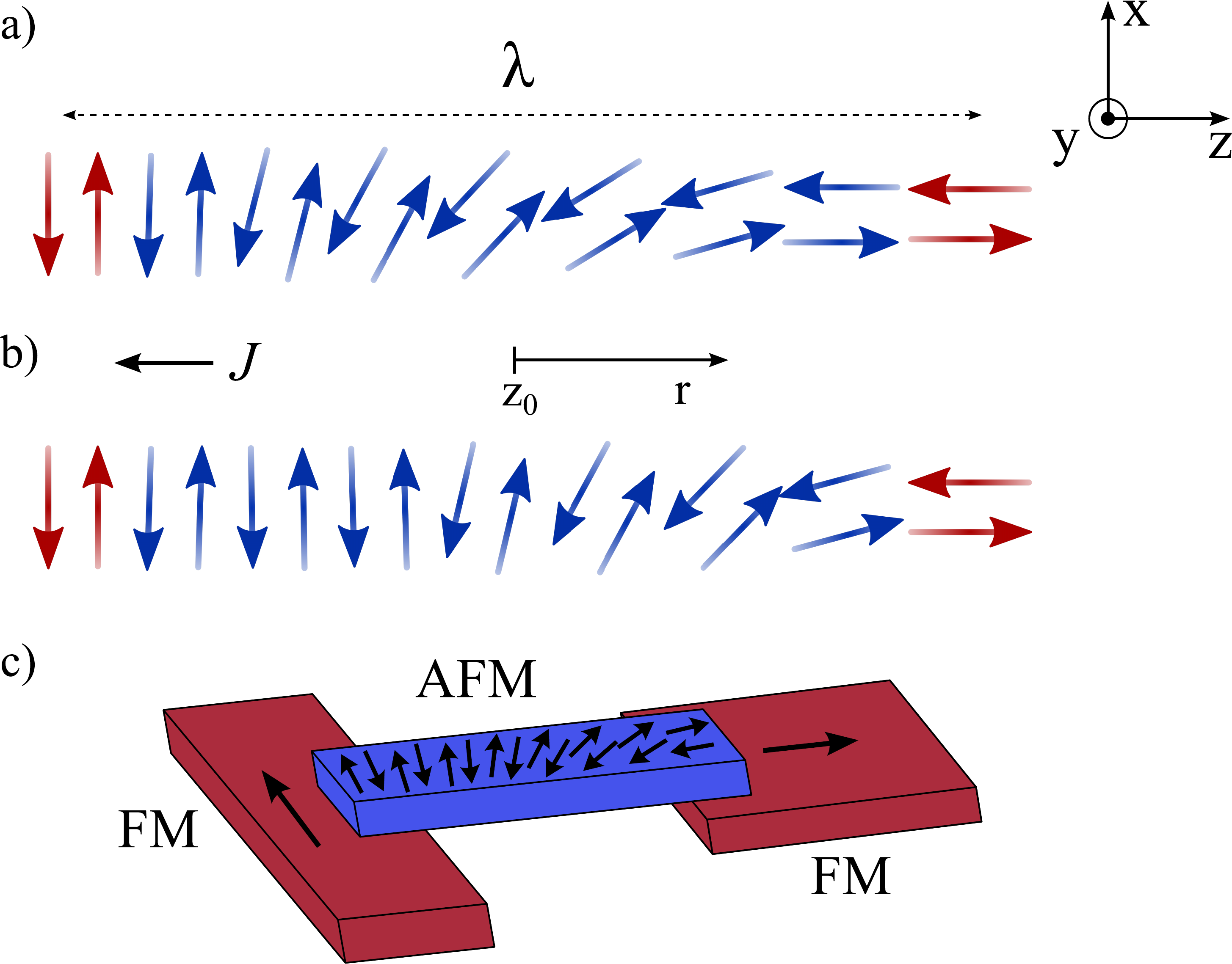}
        \caption{(color online) A 1D AFM texture pinned at a relative angle of 90$^{\circ}$ in the left and right reservoirs.
        (\textbf{a}) shows the equilibrium orientation of the staggered field, (\textbf{b}) depicts how a current $J$ exerts a torque on the staggered field vector, forcing the center coordinate $z_{0}$ to be displaced by $r$, and (\textbf{c}) shows schematically a setup of an AFM between two pinning ferromagnets.}
        \label{fig:model}
    \end{figure}

We now apply the general collective coordinate description, Eq.~(\ref{eq:eqmotion}), to an isotropic one-dimensional antiferromagnetic texture, an orientational domain wall~\cite{Bode:2006uq,*PhysRevLett.82.1020,*PhysRevLett.85.2597}, in which the antiferromagnet is pinned in the $x$ and $z$ directions at $z=0$ and $z=\lambda$, respectively. 
The staggered field $\vec{n}(z,t)$ varies slowly in the $z$ direction, see Fig.~\ref{fig:model}.
The pinning can be achieved by placing the antiferromagnet in contact with ferromagnets, as schematically shown in Fig.~\ref{fig:model}(c).
In general, the staggered field can be expressed in terms of two angles $\theta$ and $\phi$:
$\vec{n}(z,t)=\{\cos{\theta}\cos{\phi},\cos{\theta}\sin{\phi},\sin{\theta}\}$.
In equilibrium $\theta=\theta_{eq}$ and $\phi=0$, with $\theta_{eq}(z)=\pi z/(2\lambda)$.
Without loss of generality, we assume that the out-of-plane angle $\phi$ remains zero when a current passes through the system, which gives an AFM texture varying in the $x$-$z$ plane only.

In the steady state regime with a constant current along the $z$ direction, $\vec{J}=J\hat{z}$, the solution of Eq.~(\ref{eq:neeldynamics}) is $\theta_{s}(z)=\frac{\pi}{2}(1-e^{Qz})/(1-e^{Q\lambda})$, where $Q=\beta J/A$.
As a collective coordinate representing the softest mode, we use the deviation of  the texture center, $r$, from its equilibrium position $z_{0}$, which is the point where the $x$ component of the staggered field vector equals the $z$ component, $\theta (r) =\pi/4$.
In equilibrium, when there are neither applied currents nor external fields, the center coordinate is $z_{0}=\lambda/2$.
Motivated by the steady state solution $\theta_{s}$, expanding for small $Q$ in the low current regime to the linear order in the deviation $r$ from equilibrium, we use Eq.~(\ref{eq:eqmotion}) with the ansatz that the staggered field can be fully described by the sine and cosine of a function $\theta(z,r)$:
\begin{equation}
\label{eq:theta} \theta (z,r)=\frac{\pi
z}{2\lambda}\left[1+\frac{4(z-\lambda) r}{\lambda^{2}}\right].
\end{equation}
Using this ansatz and the equation of motion (\ref{eq:eqmotion}), we find that the deviation from equilibrium, $r$, obeys
\begin{eqnarray}
\label{eq:drivenoscillator} M\ddot{r} + \Gamma\dot{r}+M\omega_{0}^{2} r = F_{J}+F_{H},
\end{eqnarray}
where $M=\lambda/(a\gamma^{2})$ is the effective mass,  $\omega_{0}=\gamma(10Aa)^{1/2}/\lambda$ is the natural frequency of the system, and $\Gamma=\lambda G_{2}/\gamma$ is the damping coefficient.
There are two contributions to the external forces: 
One from the current, $F_{J}=-5\lambda[\beta J + (\eta+\beta G_{1})\dot{J}/(a\gamma)]/4$, and the other from time-varying external fields, $F_{H}=5\lambda^{2} \dot{H}_{y}/(2\pi a \gamma)$.
For DC currents, the reactive (adiabatic) force parametrized by $\eta$ plays no role, and only the dissipative (non-adiabatic) force parametrized by $\beta$ is important for the texture dynamics.
When the driving forces are independent of time, Eq.~(\ref{eq:drivenoscillator}) describes damped harmonic oscillations about a new perturbed position $r_{new}=-\beta J \lambda^{2}/(8A)$.
This solution is valid as long as $r_{new}\ll \lambda/2$.
Hence, using $\beta^{*}=\beta J d/A=-0.005$, the approach works well for systems with lengths up to several hundred lattice constants.

Numerical values for the natural frequency can be estimated for AFM metals.
For example, in FeMn, the inhomogeneous exchange constant is $A=0.94\cdot 10^{-14}$ J/m~\cite{PhysRevLett.100.226602}, the lattice constant is $d_{\rm{FeMn}}=3.6\ \mathrm{\AA}$~\cite{PhysRevB.61.11569}, and the magnetic moment per sublattice is $1.65\ \mu_{B}$, with $\mu_{B}$ being the Bohr magneton, giving a natural frequency of approximately 1 GHz for a FeMn texture with a length of 100 lattice constants.

\begin{figure}[h]
        \centering
        \includegraphics[width=0.48\textwidth]{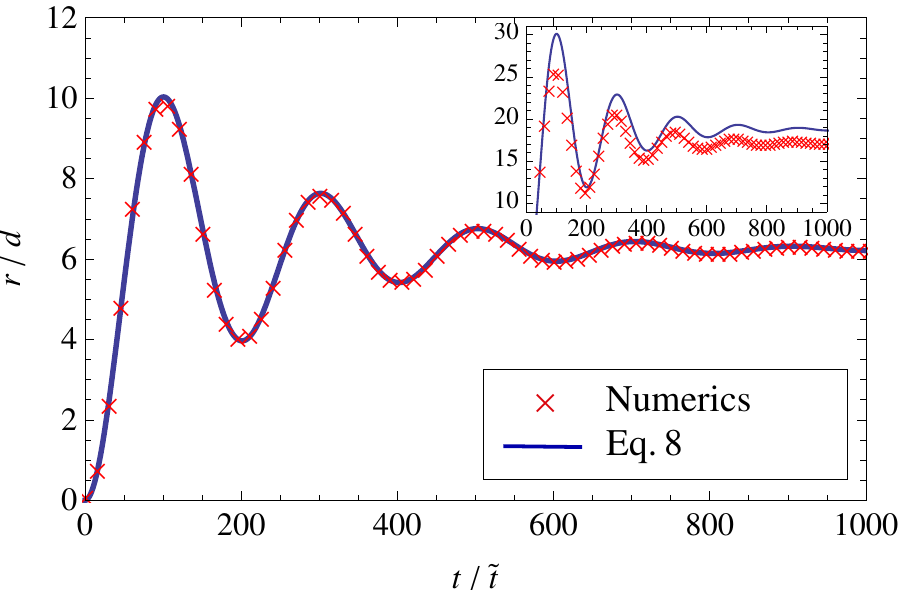}
        \caption{(color online) Transient response of the deviation $r$ from the equilibrium position $z_{0}=\lambda/2$, after a constant current has been applied at time $t=0$.
        The AFM texture shows damped oscillations around a new perturbed position.
        The magnitude of the perturbation depends on the system length and the current density. The inset shows the response when the current density is tripled.}
        \label{fig:numerics}
    \end{figure}

In Fig.~\ref{fig:numerics}, the solution of the time-dependent equation of motion for $r$, Eq.~(\ref{eq:drivenoscillator}), has been compared to numerical results of a micromagnetic simulation of the coupled Eqs.~(\ref{eq:currentmotion}) and (\ref{eq:currentmotion2}), with the boundary conditions described in Fig.~\ref{fig:model}.
The equations were first written in dimensionless form by scaling the $z$ axis with the lattice constant $d$, and the time axis with $\tilde{t}=(\gamma a l)^{-1}$. Other dimensionless quantities, as well as the numerical values used in the simulation presented in Fig.~\ref{fig:numerics}, are summarized in Table~\ref{tab:constants}.

Fig.~\ref{fig:numerics} shows that the complex spatio-temporal dynamics of the AFM texture can be described by the motion of the single soft mode $r$.
%Good agreement between the approximate collective coordinate approach and the numerical simulations is also achieved when the current density is doubled (not shown).
Fitting the simple equation of motion, Eq.~(\ref{eq:drivenoscillator}), to experimental data, e.g., from AMR measurements, can provide good estimates of the phenomenological parameters in AFMs.
The generality of Eq.~(\ref{eq:eqmotion}) also makes the collective coordinate approach a powerful tool for investigating the dynamics of more complex AFM textures with more than a single soft mode.

The staggered dynamics represented by the center coordinate $r$ can be measured via the AMR effect. The magnitude of AMR in bulk AFMs is not known, but since the tunneling AMR~\cite{Park:2011rt, *PhysRevB.81.212409} is significant, we believe its bulk value will be too.
A plausible assumption is that the simplest possible phenomenological model of AFM-AMR is similar to AMR in ferromagnets, but with the AMR depending on the orientation of the staggered field rather than the magnetization:
$\rho(\vec{n})=\rho_{0}+\rho_{ani}(\vec{n}\cdot\hat{z})^{2}$, where $\rho_{0}$ is the isotropic resistivity and $\rho_{ani}$ is the anisotropic resistivity.
Integrating the resistivity over the system, using the ansatz in Eq.~(\ref{eq:theta}), and expanding to linear order in $r$, gives the effect of the AFM texture on the resistance as $R(t)=R_{0}+\rho_{ani}[\lambda/2-8r(t)/\pi^{2}]$.
Therefore, it should be possible to observe the effects of \textit{both} DC and AC currents.
For DC currents, the total resistance $R(t)$ will be enhanced or reduced depending on the current direction.
For AC currents, by sweeping the frequency, one should observe enhanced deviations of the resistance when the frequency equals the natural frequency of the texture.
This setup offers the possibility of measuring the effect of the current-induced torque on the staggered field, a phenomenon which is, in general, difficult to observe experimentally.

\begin{table}[h!]
\centering
   \caption{Dimensionless numerical constants}
   \begin{tabular}{@{} lcc @{}}
   \hline
      Constant          	& Composition               	&   Value             	\\
      \hline
      $a^{*}$           	& $a l^{2} d^{2}/A$     		&   1                   	\\
      $\alpha_{1}$  	&   $G_{1}l$                    	& 0.01                 	\\
      $\alpha_{2}$  	&   $G_{2}/l$                   	& 0.01                 	\\
      $\beta^{*}$    	&  $\beta J d/A$           	& -0.005               	\\
      $\eta^{*}$    		&   $\eta J d/(A l)$          	& -0.005              	\\
      $\lambda^{*}$ 	& $\lambda/d$               & 100                 	\\
\hline
\end{tabular}
\label{tab:constants}
\end{table}

We can also apply our collective coordinate approach to an AFM domain wall described by the Walker ansatz: $\tan(\theta_{w})=e^{(z-r_{w})/\lambda_{w}}$.
Here we introduce the easy axis anisotropy $K_{z}$, defining the domain wall width as $\lambda_{w}=\sqrt{A/K_{z}}$. To the best of our knowledge, the experimental values of $K_{z}$ for AFM materials are still not available. However, anisotropy energies in AFMs can be comparable to, or even stronger than, those in ferromagnets since they often involve heavy elements with a strong spin-orbit interaction~\cite{umetsu:052504}.
We also reintroduce the out-of-plane tilt angle $\phi_{w}$, and use the center of the domain wall $r_{w}$,  $\phi_{w}$, and the domain wall width $\lambda_{w}$ as the three collective coordinates.
In agreement with the simplified treatment in Refs.~\onlinecite{PhysRevLett.106.107206} and \onlinecite{0953-8984-24-2-024223}, by applying a constant current, the domain wall motion gradually relaxes to a steady state, where the wall moves with the constant velocity $\dot{r}_{w}\approx-\gamma\beta J/G_{2}$.
Our approach shows that the out-of-plane tilt angle is a hard mode, which can only be excited by a time varying external magnetic field.
This is very different from the motion of domain walls in ferromagnets, where a moving domain wall also has a finite tilt angle~\cite{Tatara2008213}. 
Additionally, in linear response, there is no distortion of the domain wall width for AFMs.

In conclusion, we have derived equations of motion for the collective coordinates corresponding to soft modes of AFM textures to the linear order in currents, magnetization, and external magnetic field.
In contrast to ferromagnets, the dynamics are second order in time derivatives, e.g., the effective particles described by the soft coordinates acquire a mass, and have no first-order contribution from time-independent external magnetic fields.
We have applied our theory to a one-dimensional model of a slowly varying antiferromagnetic texture pinned at 90$^{\circ}$ at the edges, and found the natural frequency and deviations of the center coordinate in terms of the system parameters.
The results show that the dissipative (non-adiabatic) current-induced torque is crucial for the dynamics of the AFM textures.

\textit{Acknowledgments.}- This work was supported by EU-ICT-7 contract No. 257159 "MACALO". 
O.A.T. also acknowledges support from NSF under Grants No. DMR-0757992 and ONR-N000141110780.

%----------------------------------------------%
%                                                   %
%                  References                   %
%                                                   %
%----------------------------------------------%

%merlin.mbs 2010-03-15 4.21a (PWD, AO, DPC)

\end{document}